# EXPLORING THE IMPACT OF COVID-19 IN THE SUSTAINABILITY OF AIRBNB BUSINESS MODEL

## RIM KROUK[1], FERNANDO ALMEIDA[2]

[1]IAE Montpellier, France

kroukrim@gmail.com

[2]Polytechnic Higher Institute of Gaya, Portugal

almd@fe.up.pt

**Abstract:** *Society is undergoing many transformations and faces economic crises, environmental, social, and public health issues. At the same time, the Internet, mobile communications, cloud technologies, and social networks are growing rapidly and fostering the digitalization processes of business and society. It is in this context that the shared economy has assumed itself as a new social and economic system based on the sharing of resources and has allowed the emergence of innovative businesses like Airbnb. However, COVID-19 has challenged this business model in the face of restrictions imposed in the tourism sector. Its consequences are not exclusively short-term and may also call into question the sustainability of Airbnb. In this sense, this study aims to explore the sustainability of the Airbnb business model considering two theories which advocate that hosts can cover the short-term financial effects, while another defends a paradigm shift in the demand for long-term accommodations to ensure greater stability for hosts.*

## 1. INTRODUCTION

Consumer habits have experienced several significant changes over the last few years and many of these changes are mainly due to the emergence of online and cloud applications available to consumers. The tourism sector has been one of the areas where these transformations have become more visible and have allowed stakeholders in this area to see and experience the tourism experience in a different way. Several authors such as Batinic (2013), Hughes & Moscardo (2019), and Madhukar & Sharma (2019) emphasize that the increased use and importance of new digital technologies has created challenges and new opportunities for the tourism industry. Tourists are more engaged and informed through the Internet. They use several mobile applications and have access, in real-time, to news and online information about destinations, accommodation, or transportation (Dorcic et al., 2019; Law et al., 2018). Furthermore, the Internet also facilitates the sharing of experiences between users, which makes the user not only a consumer of information but also assumes a relevant role in its production (Lammi & Pantzar, 2019).

With the emergence and diversification of the online platform market, the concepts of shared economy and collaborative consumption also arose. Shared economy is a new economic model based on collaborative consumption and activities of sharing, exchanging, and renting goods. According to Wirtz et al. (2019), its essence lies in peer-to-peer (P2P) type transactions, from person to person, and in leveraging resources with an emphasis on their use, not their ownership. Frenken & Schor (2017) consider that for a company to belong to the shared economy it must have the following characteristics: (i) the business should be focused on access to underutilized goods; (ii) consumers should benefit from access to goods and services; and (iii) the business should be supported by decentralized networks and marketplaces. May et al. (2017) share the same vision in stating that one of the characteristics of the shared economy is to make profitable assets that are little used and, in this way, to take advantage of market gaps.

According to the actual sanitary conditions, in the context of a worldwide pandemic, a lot of companies need to make crucial decisions. Either close or choose a way to adapt their business. Some of them were really innovative and found creative ways to survive an unknown situation posed by COVID-19 (Almeida, 2020; Harris et al., 2020). The Internet was a major way to maintain the sale for some, but for other enterprises that furnish services, it is time to innovate and find ways to adapt their offer in order to take advantage of the situation. Flexibility and agility emerge as two crucial factors for the tourism industry to adapt and mitigate the effects of travel restrictions (Ugur & Albiyik, 2020). The travel agencies, airlines, hotels, seasonal infrastructures, and more were the first ones to take the blow of the crisis. The World Tourism Organization estimates that revenues from tourism could fall by $910 billion to $1.2 trillion in 2020 and between 100 to 120 million direct tourism jobs are at risk (UNWTO, 2020).

Some succeeding economic models based on the shared economy were highly impacted too. In this paper, we're going to focus on the service provider platform, Airbnb, in order to highlight the extent of the crisis. Since its establishment in 2008, Airbnb has become one of the largest online accommodation providers and an epitome of the shared economy. The firm has gained public and scholarly attention due to its disruptive effects on the hospitality industry, impacts on housing markets, and legal conflicts over housing, taxation, and consumer regulations. Today, the CEO of Airbnb, Brian Chesky, qualified these times as "the most painful crisis of our lives" (Airbnb, 2020).

This paper will seek to understand and explore how Airbnb, as well as similar companies using the same economic model, can survive considering the context of this global pandemic. Equally relevant is exploring the innovative decisions that can be made to ensure their sustainability in this period. This manuscript is organized as follows: In the first phase, a review is made of the literature on the shared economy and the role of Airbnb in this business model. Then, in the methodology, the methods adopted for information collection, analysis, and interpretation are presented. After that, the results are discussed considering Airbnb's evolution and sustainability perspectives. Finally, the main conclusions of the study are presented, and some indications of future work are given.

## 2. LITERATURE REVIEW

Shared economy and collaborative consumption are concepts that are incorporated in a business model characterized by the supply of goods, services or use of resources in a P2P model, without any associated acquisition or purchase. From a conceptual perspective, the shared economy represents a model based on the use of a good or service, without the acquisition of the same, through a form of temporary ownership (Mont et al., 2020). Therefore, this model replaces the classic concept of ownership. Although sharing practices have always existed, which makes the shared economy innovative is the constant exchange between unknown people, which is only possible due to technological development (Sutherland & Jarrahi, 2018). This model replaces the old sharing practices that were only carried out between people in their proximity.

Technological advances and economic factors may appear to be the main factors that have served as a pillar for the growth of collaborative consumption and the shared economy. However, they are not the only determining factors. Paulauskaite et al. (2017) point out that the shared economy and collaborative consumption have also influenced the way people experience, consume, and produce tourism products. The transactions among consumers have opened up a greater possibility for tourists to have a more authentic experience and feel the comfort of a household outside the home. For many tourists, a social experience with a more domestic side can be a very important factor for the use of shared economy accommodation platforms, since this type of experience can hardly be acquired in the more traditional tourism industry (Pung et al., 2020).

The available literature in the shared economy allows us to identify three main areas of research. Firstly, there is research related to the business model that sustains this type of economy and its effects on the various players (Barbu et al., 2018; Ritter & Schanz, 2018); secondly, there are researchers looking at the social phenomenon of the shared economy (Curtis et al., 2020; Habibi, 2019); and finally, the third dimension explores sustainable development and the collaborative consumption economy (Liu & Chen, 2020; Mi & Coffman, 2019). According to Kim (2019), the most decisive factors stimulating the development of the shared economy are the economic ones (e.g., consumer behavior, increasing population density, global recession). However, other factors should also not be forgotten like the technological and environmental dimensions (Pouri & Hilty, 2018; Wu & Zhi, 2016).

A lot of firms started to use this economic model such as Uber and Airbnb for example. The sharing economy (SE) has allowed travelers to organize their trips in new ways. They can book a room through Airbnb, travel in a car booked on Uber, eat food made by locals with EatWith and move around a city with a shared bike, all at a lower price than they would conventionally pay (Heo, 2016). These startups, like Airbnb and Uber, have experienced enormous growth in recent years and operate on a global scale. This makes them face and compete directly with more conventional hotel and transport companies and, in most cases, they are more effective than their traditional competitors, namely in a competition based on a lower price that makes them more appealing to the consumer (Penn & Wihbey, 2016; Hoffman, 2020).

We are currently observing a growing phenomenon of dematerialization in Western societies whose pace has accelerated strongly in the last year in view of the processes of confinement and increase of digital transactions between companies and people caused by COVID-19 (Almeida et al., 2020). In general, it can be concluded that dematerialization refers to a relative reduction in the amount of materials or physical goods needed to perform economic functions. Hadad & Bratianu (2019) refer that these dematerialization phenomena gain greater expression in the digital sphere and in the intangible aspects of consumption. As a result of this process, new organizational models such as Airbnb have emerged, which has fostered the emergence of local temporary housing. Airbnb is one of the most relevant collaborative consumer companies in the panorama and Guttentag (2015) characterizes it as a disruptive innovation because it has decisively influenced the temporary accommodation market and even the way of doing tourism. Airbnb can be used by tourists, travelers, and professionals and allows the rental of a property without much bureaucracy in a simple way and with lower costs (Guttentag et al., 2018; Steward, 2019). The whole process of booking accommodations, payment, message exchanges takes place within the platform itself.

Airbnb's business model still has few regulatory mechanisms when compared to those facing hotel accommodation. However, some cities (e.g., New York, Los Angeles, Boston) have proposed legal mechanisms to limit the possibility of having paying guests for short periods of time and special licenses are required to advertise a property and receive guests (Hamed, 2019). The way these restrictions are implemented depends heavily on each local government and does not apply only in the USA. Also, in Europe, limitations on Airbnb's operation can be found in Berlin and Barcelona (Doward, 2016). This government supervision is intended to control the number of tourists in each city and to stimulate the rental market to fight the lack of affordable housing for residents.

Studies are also emerging that show that the offer made available by Airbnb has negative impacts on local hotel revenues. According to Guttentag et al. (2018), its negative impacts are mainly on hotels of a lower range. Business owners of the hotel sector point out that the company's tax burden tends to be lower because it is only an application that makes the connection between advertisers and travelers which allows the company to offer lower prices (Bivens, 2019). This situation has caused a drop in hotel revenues as demonstrated in the study conducted by Zervas et al. (2017). However, this impact does not seem to be common to all destinations. The study developed by Blal et al. (2018) revealed that the volume of the Airbnb offer does not influence the growth trajectory of local accommodation establishments since Airbnb is seen by travelers as a complementary service. Finally, Guttentag (2019) considers that the experience offered by Airbnb has reshaped travelers' expectations regarding their stay, since one of the main transformations caused by the platform was to put a greater focus on the traveler's experience.

The growing importance of the economy of sharing does not mean that a fundamental change is taking place in the social perspective and in human relations, or that it represents the end of the traditional economy. What emerges as an innovative factor are the business models that allow rapid contact between unknown people who initiate transactions to meet their needs (Ritter & Schanz, 2019). According to Wirtz et al. (2019), the rapid expansion of the shared economy was essentially due to new businesses based on peer-to-peer platforms, which made it possible to share goods and services. However, unlike Airbnb, many of these companies are startups with a low level of maturity and high risks to the development of their activity. Moreover, legal obstacles also arise because they are businesses that do not yet have formal regulation (Cauffman & Smits, 2016).

## 3. METHODOLOGY

This study adopts the qualitative methodology for the analysis of the effects of COVID-19 on the sustainability of the Airbnb business model. According to Queirós et al. (2019), qualitative research allows understanding a phenomenon and interpreting its values and relationships, not dissociating the thinking from the reality of social actors where the intervenient plays an active role in the development of new knowledge. This type of research aims to understand a phenomenon based on its observation and contextualization. Therefore, the researcher seeks to understand the importance of the phenomena studied, according to the perspective of the participants of the situation studied and then the interpretation of the phenomena studied is done (Yin, 2015). In the context of the challenges posed by COVID-19, the qualitative methodology is indicated to allow understanding a phenomenon with badly defined boundaries and limits and whose impacts have to be accessed from multiple perspectives.

Figure 1 gives a concise overview of the adopted methodology phases. It is organized into three phases: (i) contextual stage; (ii) design stage; and (iii) analysis stage. In the contextual dimension, the aim is to contextualize the impact of COVID-19 on the tourism sector, considering that this sector was particularly affected by the pandemic. Still, at this stage, the characteristics of the shared economy are presented and how it becomes relevant in the context of globalization. The Airbnb business model is also analyzed. Then, in the design stage, the research methods that support these studies are presented. Still, in this stage, the sources of information are presented. and the process of data collection and interpretation are described. Finally, in the analysis stage, theories supporting the future evolution of Airbnb are presented considering the impact of COVID-19 on company activities and new social trends and consumer behaviors resulting from the pandemic. It is also at this stage that the main implications of this study are addressed and some perspectives of future research in the area are given.

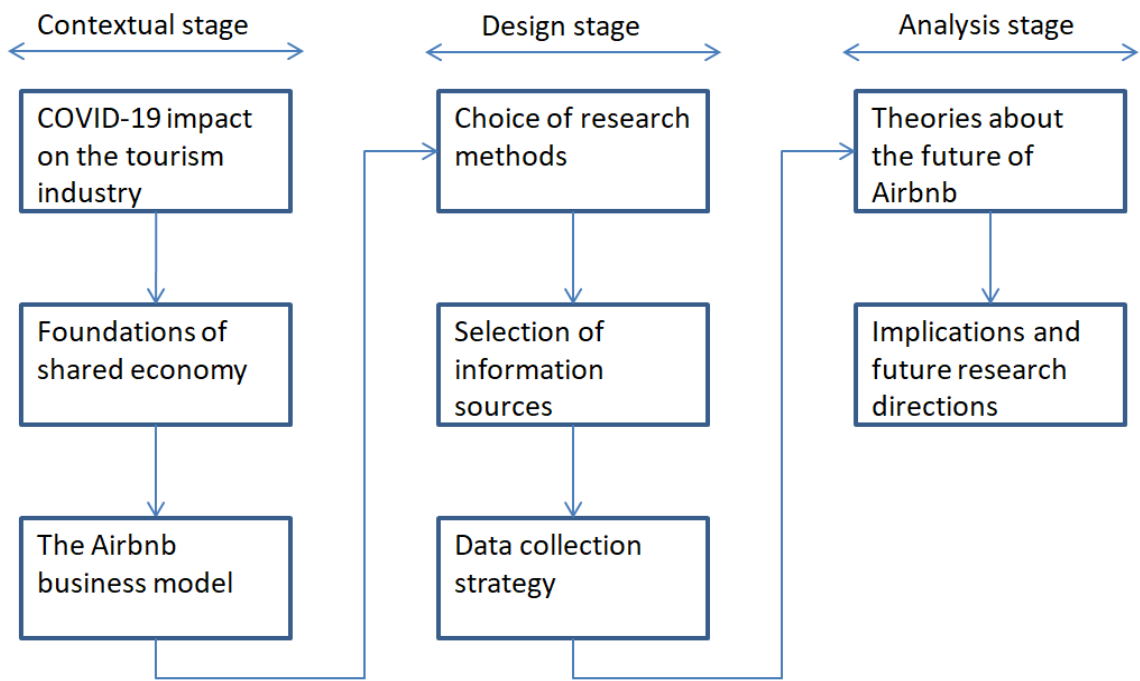

Fig. 1 – Phases of the adopted methodology

Secondary sources of information were used in the data collection process. Five types of information were considered, specifically: (i) Airbnb's institutional website that allows understanding the actions taken by the company to face the limitations imposed on the movement of people and hygiene measures; (ii) books on the business model published by Hoffman (2020) and Steward (2019) that seek to address the role of short-term rental and Airbnb's operating process; (iii) research published in scientific journals and international conferences that allows exploring the impact of COVID-19 on the sustainability of Airbnb's operations; (iv) customer and host reviews that allow identifying the main concerns of the agents involved in the process; and (v) personal and corporate blog that are also relevant to have a sufficiently comprehensive perspective on the various paradigms of Airbnb's evolution in the market. The adoption of multiple sources of information, simultaneously diversified and comprehensive, seeks to capture the various perspectives and contribute to increasing the robustness of the investigation process.

## 4. ANALYSIS AND DISCUSSION OF RESULTS

Even though the COVID-19 has affected all sectors, it has particularly affected companies the shared economy model in the touristic sector. The simultaneous combination of these two factors makes companies very vulnerable to the effects of COVID-19 in areas such as income reduction, job loss, and anxiety (Hossain, 2020). Above all, service providers are among the most vulnerable. Airbnb lost in a few weeks 25% of its workforce, or about 1,900 people. Furthermore, significant cuts were imposed on the marketing area and other sectors (Syed, 2020). Additionally, because it's really unclear if Airbnb hosts should be considered employees, contractors, or workers. Airbnb considers their hosts as contractors, and so they do not have the responsibility for insurance and social benefits (Farmaki et al., 2020).

In the case of Airbnb, many hosts use their activity on the platform as their full-time occupation and have multiple properties that are rented from property owners, so they are in highly vulnerable situations. Due to the COVID-19, the occupancy rate of properties was zero in many cities from one to more than three months. It is important to mention that Airbnb decided to adopt a full refund policy for the guests at the beginning of the corona crisis because of the political travel restrictions imposed by the governments. However, many hosts believe that guests cancelled their bookings to receive a full refund, even if their reasons for cancelling were unrelated to the pandemic, and they feel that Airbnb did not consult with them in the decision-making process. Consequently, some hosts had no income and felt abandoned by Airbnb (George et al., 2020).

The question now is to know how Airbnb is going to react facing COVID-19 and all it implies. They already lifted a $250m fund to support with cancellations caused by the pandemic and Members of the Superhost club, which gives Airbnb hosts "more visibility, earning potential, and exclusive rewards", also have access to a $17m fund to help pay the rents of their listed properties. (Temperton, 2020). In order to implement these measures, the company had to make a cut in its budget, which caused a layoff of about a quarter of their direct employees (Hu & Lee, 2020). Airbnb currently recommends that hosts adopt a more flexible cancellation policy and make their calendar available for longer stays with weekly or monthly discounts. The goal of adopting a more flexible cancellation policy is to increase guests' confidence in the market so that they can proceed with a new reservation under conditions of great instability in the area of public health; while having a more comprehensive calendar demands that the reservations made can be longer and give hosts greater predictability about the market demand. Other more innovative initiatives emerged as the concept of online experiences that allowed Airbnb to offer unique online experiences about the cultures and traditions of each region, such as cooking classes or guided mediations (Freedman, 2020). With this, the aim is to offer a unique entertainment experience to the tourist without having to leave home, but that allows the tourist to know the playful, social, and historical context of each place. In addition to meeting a short-term need, these initiatives also seek to arouse in the tourist the desire to know these places when the measures to restrict the movement of people have been lifted.

### 4.1 Opposing theories about the future of Airbnb

It is possible to predict the future of Airbnb considering the multiple visions of hosts and guests' reactions and on how the company is going to adapt its business model to face the pandemic and everything it implies. Two hypotheses can be formulated:

- Hypothesis I: Hosts who are able to afford to cover their expenses for the duration of the crisis, may choose to remain in the short-term market (would help the sustainability of Airbnb);

- Hypotheses II: Hosts who are not able to afford short-term expenses will opt for mid-term and long-term rentals due to their relative stability to ensure their expenses will be covered (would not help the sustainability of Airbnb).

### 4.1.1 Hypothesis I

Two kinds of hosts exist: the ones that use Airbnb as a subletting business based on rental-arbitrage with several listings and the others who are renting their principal residence. In both cases, there are many advantages for them and, accordingly, they will not leave the platform and continue short-term rentals.

Both types want by all means to avoid stable leases because it implies a lot more risk and less profitability. Accordingly, a transfer of homes to long-term rentals will probably not occur as a general trend. Moreover, the short-term rental market is way more highly profitable and practical for landlords. They can sell the property, use it, or directly abandon the market when they will with no legal procedure concerning the end of a lease (since there's none). Another advantage of this control over the asset is that the maintenance of the property is constant since it is possible to check the condition of the apartments weekly; that tourists pay in advance; and that the high turnover of "tenants" allows landlords to constantly speculate with rental prices (Cocola-Gant, 2020).

Unfortunately, all these businesses and the rise of commercial intermediaries managing properties for third parties, suggests that Airbnb was shifting away from its original philosophy, becoming primarily a commercial space trading platform. We hypothesize that the COVID19-induced super-shock will lead to a re-emergence of the original Airbnb ethos (O'Neill & Ouyang, 2016).

An argument is also important to understand why hosts would want to continue renting short-term. When the touristic activities will start again or even for the local or business travelers, booking an Airbnb might become the safest accommodation the tourist can find. Touristic apartments, unlike hotels, might have many years ahead of them, allowing the customer to isolate himself and feel complete safe. Maybe travelers would feel more secure in private condos rather than in a hotel resort where they will find themselves interacting with more people (AFP, 2020). This perspective is confirmed in the study performed by Cheng et al. (2020) that intended to estimate the impact of COVID-19 on the sharing economy. The findings of this study reveal that entire homes were more booked than other kinds of accommodations. However, this doesn't mean that entire homes are the perfect kinds of accommodations in all minds. As reveal by Hawlitschek et al. (2016) and Räisänen et al. (2020), trust between users is one of the most important factors in the shared economy model. COVID-19 has created mistrust between customers and hosts beyond the transaction stage (Yang et al., 2019). Therefore, it is expected that firms will experience difficulties exercising control over safety rules and protocols to minimize coronavirus spread. There're actually two ways of seeing it for the problem of sanitary conditions: (i) people may now favor traditional hotels over home-sharing because of hygiene standard because it is difficult to guarantee a deep clean on a host-to-host basis after every guest; and (ii) people will prefer service apartments over sharing accommodation and hotels due to hygiene and social distancing, even after the COVID-19 crisis. Currently, Chua et al. (2019) state that the real challenge for Airbnb will be to maintain the level of trust of its consumer to ensure its sustainable growth.

The understanding of the Airbnb sustainability phenomenon is also strongly related to consumption habits and the role of transport companies, especially in low-cost airline transport which has been a key pillar for the growth of Airbnb (Kunwar, 2020). This has allowed the capture of new audiences that were not used to traveling, or that traveled less frequently, and that now have much greater and diversified access. However, COVID-19 challenged this paradigm. The challenges are not only short-term but may also have long-term repercussions. Therefore, the changes we are experiencing in the market may not be temporary or of short duration. As Sheth (2020) points out, COVID-19 may have long-term effects on consumers' consumption patterns. Also, in the tourism field, people are expected to seek accommodation closer to home, safer, and more affordable.

### 4.1.2 Hypothesis II

The COVID-19 has prompted firms and service providers to think about their services differently, and many have adopted strategies to mitigate the effects of the COVID-19. Accommodation hosts are considering finding mid or long-term tenants and focusing on domestic rather than foreign guests. This situation occurred in the summer of 2020 when several European countries saw significant growth in the domestic tourism market. For example, Portugal saw a doubling in this period in the demand for internal reservations in spaces that were not as usual as spacious houses to house families, equipment to receive pets, or longer stays (Antunes, 2020). It is expected that once restrictions will be lifted demand for Airbnb-listed properties will increase again. But not all hosts will return to the short-term market. Hosts will now factor into their calculations the risk associated with economic super-shocks. Supply will reach an upper limit, irrespective of the demand (Dolnicar & Zare, 2020). Landlords using short-term rentals may have realized how dependent they were on their environment. The problem is that with the actual restrictions they have no power on the reservation. There are no more international tourists which were the main source of income usually. And with no guests, there's nobody to pay the rent, and that leaves Airbnb entrepreneurs open to huge financial liabilities (Tenderson, 2020). This is the reason why some of them might add flexibility to their business and move to other forms of rental during times when the short-term is no longer profitable.

Many cities worldwide are still struggling to find ways to regulate Airbnb (Guttentag, 2015). In general, three regulatory approaches have been identified in the existing literature: prohibition, laissez-faire, and allowing it with certain restrictions (Jefferson-Jones, 2014). Since some governments such as Spain, are allowing "seasonal leases" it's easier to avoid making long-term contracts and keep their advantages by going to mid-term. The change from short- to mid-term rentals is what seems to be happening during the current pandemic. It's a bit similar to 'seasonal leases' which propose the property for students (mainly) from September to June and practice short-term during summer. But here, landlords will switch way more often from one type of renting to another because the situation is too uncertain for now.

## 5. CONCLUSION

Shared economy has the potential to revolutionize the way we buy and sell, use and provide services, and think about market transactions. Shared economy and collaborative consumption businesses can generate change especially in the tourism sector, as they attract new profiles of travelers as individuals looking for low prices, local experiences, or environmental concerns. Furthermore, these platforms are also a means of facilitating distinctive and authentic social encounters between tourists and locals and contribute to the reduction of accommodation and transportation prices for travelers.

Shared economy platforms are highly vast, but some platforms stand out for their success and the way they have changed the paradigm of consuming certain services or products. Airbnb is one of the best-known platforms that has changed the paradigm in the accommodation sector. However, it was also one of the platforms that was most affected by the COVID-19 pandemic, due to its model based on the shared economy in the touristic sector. These effects had a direct impact on the reduction of revenue and job losses, with simultaneous impacts on both the economic and social dimensions. The company took short-term measures related to the cancellation policy, security and cleanliness, long-term reservations, and online experiences.

These measures, despite their relevance, are focused on an immediate response to the challenges posed by COVID-19 to the operation of the business. However, their effects are much deeper, and should be analyzed from the perspective of their sustainability.  At this level, two hypotheses were discussed: (i) maintenance of Airbnb's business model considering that the financial effects of the pandemic are covered by the hosts; and (ii) change in the behavior of the hosts that can focus on a longer lease that allows them to ensure greater stability.

This study offers both theoretical and practical contributions. From a conceptual perspective, it was possible to deepen the knowledge about the shared economy and the impact of COVID-19 on this business model. In the practical dimension, this study is relevant especially for Airbnb hosts to realize what strategies they can adopt to ensure the sustainability of their business model while the effects of the pandemic are being felt. Predicting tourist behavior in the shared economy is essential for making informed decisions that will both address the sharp reduction in demand in the short term and ensure the sustainability of operations in the long term. As future work, it becomes relevant to explore the impact of COVID-19 on other businesses in the shared economy that are not restricted to the accommodation sector. It is also relevant to explore the effects of COVID-19 on the economy of some cities that have grown strongly in recent years due to the proliferation of local housing and low-cost airlines.

**CONFLICTS OF INTEREST AND PLAGIARISM:** The authors declare no conflict of interest and plagiarism.